\documentclass[8pt]{article}
\usepackage{amsmath,amssymb}

\makeatletter
\@addtoreset{equation}{section}

\makeatother

\def\un#1{\relax\ifmmode\@@underline#1\else
        $\@@underline{\hbox{#1}}$\relax\fi}

\let\du=\du                     % dot-under
\def\a{\alpha}
\def\b{\beta}

\def\d{\delta}

\def\h{\eta}

\def\k{\kappa}

\def\r{\rho}

\def\bo{{\raise-.3ex\hbox{\large$\Box$}}}               % D'Alembertian
\def\TH{{\raise.2ex\hbox{$\displaystyle \bigodot$}\mskip-4.7mu \llap H \;}}
\def\face{{\raise.2ex\hbox{$\displaystyle \bigodot$}\mskip-2.2mu \llap {$\ddot
        \smile$}}}                                      % happy face
\def\leftrightarrowfill{$\mathsurround=0pt \mathord\leftarrow \mkern-6mu
        \cleaders\hbox{$\mkern-2mu \mathord- \mkern-2mu$}\hfill
        \mkern-6mu \mathord\rightarrow$}
\def\dvec#1{\vbox{\ialign{##\crcr
        \leftrightarrowfill\crcr\noalign{\kern-1pt\nointerlineskip}
        $\hfil\displaystyle{#1}\hfil$\crcr}}}           % <--> accent
\def\dt#1{{\buildrel {\hbox{\LARGE .}} \over {#1}}}     % dot-over for sp/sb
\def\frac#1#2{{\textstyle{#1\over\vphantom2\smash{\raise.20ex
        \hbox{$\scriptstyle{#2}$}}}}}                   % fraction
\def\sfrac#1#2{{\vphantom1\smash{\lower.5ex\hbox{\small$#1$}}\over
        \vphantom1\smash{\raise.4ex\hbox{\small$#2$}}}} % alternate fraction
\def\bfrac#1#2{{\vphantom1\smash{\lower.5ex\hbox{$#1$}}\over
        \vphantom1\smash{\raise.3ex\hbox{$#2$}}}}       % "
\def\afrac#1#2{{\vphantom1\smash{\lower.5ex\hbox{$#1$}}\over#2}}    % "
\def\[{\lfloor{\hskip 0.35pt}\!\!\!\lceil}
\def\]{\rfloor{\hskip 0.35pt}\!\!\!\rceil}

\def\du#1#2{_{#1}{}^{#2}}
\def\ud#1#2{^{#1}{}_{#2}}
\def\dud#1#2#3{_{#1}{}^{#2}{}_{#3}}
\def\udu#1#2#3{^{#1}{}_{#2}{}^{#3}}

\def\fracm#1#2{\hbox{\large{${\frac{{#1}}{{#2}}}$}}}
\def\ha{{\fracmm12}}

\def\un{\underline}
\def\fracmm#1#2{{{#1}\over{#2}}}

\def\low#1{{\raise -3pt\hbox{${\hskip 0.75pt}\!_{#1}$}}}

\def\Dot#1{\buildrel{_{_{\hskip 0.01in}\bullet}}\over{#1}}
\def\dt#1{\Dot{#1}}
\def\DDot#1{\buildrel{_{_{\hskip 0.01in}\bullet\bullet}}\over{#1}}
\def\ddt#1{\DDot{#1}}

\def\DDDot#1{\buildrel{_{_{\hskip 0.01in}\bullet\bullet\bullet}}\over{#1}}
\def\dddt#1{\DDDot{#1}}

\def\DDDDot#1{\buildrel{_{_{\hskip 
0.01in}\bullet\bullet\bullet\bullet}}\over{#1}}
\def\ddddt#1{\DDDDot{#1}}

\newskip\humongous \humongous=0pt plus 1000pt minus 1000pt

\newif\ifdtup

\newcommand{\be}{\begin{equation}}
\newcommand{\ee}{\end{equation}}
\newcommand{\nbe}{\begin{equation*}}
\newcommand{\nee}{\end{equation*}}

\newcommand{\lb}{\label}

\renewcommand{\[}{\left[}
\renewcommand{\]}{\right]}

\begin{document}

\thispagestyle{empty}

\noindent
\vskip2.0cm
\begin{center}

{\large\bf ON THE QUARTIC CURVATURE GRAVITY \vglue.1in 
           IN THE CONTEXT OF FRW COSMOLOGY~\footnote{Supported in part 
by the Japanese Society for Promotion of Science (JSPS)}}

\vglue.3in

Masao Iihoshi~\footnote{Email address: iihosi-masao@ed.tmu.ac.jp}
and Sergei V. Ketov~\footnote{Email address: ketov@phys.metro-u.ac.jp}

\vglue.1in
{\it Department of Physics\\
     Tokyo Metropolitan University\\
     1--1 Minami-osawa, Hachioji-shi\\
     Tokyo 192--0397, Japan}
\end{center}
\vglue.1in
\begin{center}
{\Large\bf Abstract}
\end{center}
\vglue.1in
\noindent We consider the purely gravitational fourth-order (in the spacetime 
curvature) quantum corrections to the Einstein-Hilbert gravity
 action, coming from superstrings in the leading order with respect to
the Regge slope parameter, and study their impact on the evolution of the 
Hubble scale in the context of the FRW cosmology, in four spacetime dimensions.
We propose the generalized Friedmann equations, and rule out the most naive
 (Bel-Robinson tensor squared) gravity. Our new cosmological equations have 
exact inflationary solutions without a spacetime singularity. 

\section{Introduction}

The homogeneity and isotropy of the Universe, as well as the observed spectrum
of density perturbations, are explained by inflationary cosmology \cite{inf}.
Inflation is usually realised by introducing a scalar field (inflaton) and 
choosing an appropriate scalar potential. By using Einstein equations it gives
 rise to a massive violation of the strong energy condition, and the need of an
 exotic matter with large negative pressure (dark energy). Despite of the  
simplicity of many inflationary scenarios, the origin of their key ingredients,
 such as the inflaton and its scalar potential, remain obscure. As is well 
known, the Standard Model of elementary particles has no inflaton.

Being the strongest candidate for a unified theory of Nature, including a 
consistent quantum gravity sector, the superstrings/M-theory provide the 
natural arena for building specific mechanisms of inflation. In recent years, 
many brane inflation scenarios were proposed (see e.g. ref.~\cite{binf} for a 
review), including their embeddings into the (warped) compactified superstring 
models, in a good package with the phenomenological constraints coming from 
particle physics (see e.g. ref.~\cite{kklt}). However, it did not contribute to
 revealing the orgin of the key ingredients of inflation. It also greatly 
increased the number of possibilities up to $10^{100}$ (known as the String 
Landscape), hampering any specific theoretical predictions in the search for 
signatures of strings and branes in the Universe. 

The inflaton driven by a scalar potential, or their engineering by strings and 
branes, are by no means required. The alternative could be a modification of 
the gravitational part of Einstein equations by terms of the higher order in 
the spacetime curvature \cite{starob}. It needs neither an inflaton nor an 
exotic matter, while the higher-curvature terms do appear in the effective
action of superstrings \cite{book}.

The perturbative strings are merely defined on-shell (in the form of quantum 
amplitudes), while they give rise to the infinitely many higher-curvature 
corrections to the Einstein equations, to all orders in the Regge slope 
parameter $\a'$ and the string coupling $g_s$, whose finite form is unknown and
is beyond our control. However, it still makes sense to consider the leading 
corrections to the Einstein equations, coming from strings and branes. Being 
valid for limited energy scales, the results to be obtained from them cannot 
be conclusive, but they may offer both qualitative and technical insights into 
cosmology, within the well defined and very restrictive framework.  We 
reconsider the fundamentals of that approach towards inflation, based on the 
Einstein equations modified by the leading superstrings-generated gravitational
 terms to be treated on equal footing with the Einstein terms, 
i.e. non-perturbatively.

We consider only geometrical (i.e. pure gravity) terms in the low-energy 
superstring effective action in four space-time dimensions. We assume that the 
quantum $g_s$-corrections can be suppressed against the leading 
$\a'$-corrections, whereas all the moduli, including a dilaton and an axion, 
are somehow stabilized (e.g. by fluxes, after compactification to four 
dimensions), so that the naive dimensional reduction of the quantum gravity 
terms is valid.

\section{Superstrings-modified gravity equations}

The only purely gravitational terms, coming from type-II superstrings in 
{\it four\/} spacetime dimensions, are given by
\be
S_4  =  -\fracmm{1}{2\k^2}\int d^{4}x\,\sqrt{-g}\left( R +\b J_R \right)
\lb{4a}
\ee
where $\beta$ is a new coupling constant, and
\be J_R= R^{mijn}R_{pijq}R\du{m}{rsp}R\ud{q}{rsn}
+\fracmm{1}{2}R^{mnij}R_{pqij}R\du{m}{rsp}R\ud{q}{rsn}+O(R_{mn})
\lb{oco4}
\ee
that can also be rewritten as the Bel-Robinson tensor squared \cite{br}.

As regards the four-dimensional heterotic strings, the action (\ref{4a}) is
to be supplemented by the term \cite{het2}
\be 
S_H  =  
-\fracmm{1}{2\k^2}\int d^{4}x\,\sqrt{-g}\left( \fracm{1}{8}J_H\right)
\lb{het}
\ee
where 
\be J_H = R_{ijkl}R^{ijkl} + {\cal O}(R_{mn}) \lb{oho}
 \ee
again modulo Ricci-dependent terms.
 
The gravitational action is to be added to a matter action, which lead to the
{\it modified\/} Einstein equations of motion (in the type II case, for 
definiteness)
\be
R_{ij}- \fracm{1}{2}g_{ij}R +\b \fracmm{1}{\sqrt{-g}}\fracmm{\d}{\d g^{ij}}
\left( \sqrt{-g}J_R\right) =\k^2 T_{ij} \lb{meo}
\ee
where $T_{ij}$ stands for the energy-momentum tensor of all the matter fields 
(including dilaton and axion).

There is about a hundred of the Ricci-dependent terms in the most general 
off-shell gravitational effective action {\it quartic} in the curvature. It 
also means about 100 of the new coefficients, which makes fixing the off-shell 
action to be extremely difficult. The quartic curvature terms are thus 
different from the {\it quadratic} curvature terms, present in the on-shell 
 heterotic string effective action (\ref{het}), whose off-shell extension is 
very simple (see below). 

The absence of the higher order time derivatives is usually desirable to 
prevent possible unphysical solutions to the equations of motion, as well as 
preserve the perturbative unitarity, but it is by no means necessary. As is 
well known, the standard Friedmann equation of General Relativity is an 
evolution equation, i.e. it contains only the first-order time derivatives of 
the scale factor \cite{inf,ll}. It happens due to the cancellation of terms 
with the second-order time derivatives in the mixed $00$-component of Einstein 
tensor --- see e.g. Appendix of ref.~\cite{first} for details. It can also be 
seen as the consequence of the fact that the second-order dynamical 
(Raychaudhuri) equation for the scale factor in General Relativity can be 
integrated once, by the use of the continuity equation, thus leading
 to the evolution (Friedmann) equation. As regards the quadratic 
curvature terms present in the heterotic case, their unique off-shell extension
 is given by the Gauss-Bonnet-type combination \cite{bdes}
\be 
J_H\to G = R_{ijkl}R^{ijkl} - 4R_{ij}R^{ij} + R^2 \lb{gb}
\ee 
In the expansion around Minkowski space, 
$g_{ij}(x)=\h_{ij} + h_{ij}(x)$, the fourth-order derivatives (at the 
leading order in ${\cal O}(h^2)$) coming from the first term in eq.~(\ref{gb}) 
cancel against those in the second and third terms \cite{zwie}. As a 
result, the off-shell extension (\ref{gb}) appears to be ghost-free in any 
dimensions. As regards {\it four} space-time dimensions, the terms (\ref{gb}) 
can be rewritten as the four-dimensional Euler density. Therefore, 
being a total derivative, eq.~(\ref{gb}) does not contribute to the 
four-dimensional effective action.
  
The matter equations of motion in General Relativity imply the covariant 
conservation law of the matter energy-momentum tensor,
\be (T^{ij})_{;j}=0 \lb{em}
\ee
By the well known identity $(R^{ij}-\ha g^{ij}R)_{;j}=0$, eqs.~(\ref{meo}) and 
(\ref{em}) yield
\be
\left[ \fracmm{1}{\sqrt{-g}}\fracmm{\d}{\d g^{ij}}
\left( \sqrt{-g}J\right)\right]_{;j}=0 \lb{div}
\ee
For instance, when $J=G$ as in eq.~(\ref{gb}), eq.~(\ref{div}) reads
\begin{align} 
-\ha (R_{ijkl}R^{ijkl} - 4R_{ij}R^{ij} + R^2)_{;m}+
2(R_{mjkl}R^{njkl})_{;n} \nonumber  \\
- 4(R_{minj}R^{ij})^{;n}-4(R_{mi}R^{in})_{;n}+2(RR_{mn})^{;n}  =0
\lb{div1}
\end{align}
By the use of Bianchi identities for the curvature tensor, we found that the 
left-hand-side 
of eq.~(\ref{div1}) identically vanishes in the case of Gauss-Bonnet gravity. 
Equation (\ref{div}) should be {\it identically\/} satisfied by any 
off-shell gravitational correction $J$.

Given the quartic curvature terms (\ref{oco4}), the modified Einstein equations
 of motion (\ref{meo}) are given by
\begin{align} 
\k^2 T_{ij} & = R_{ij}-\ha g_{ij}R +\b \left[  -\ha g_{ij}J_R 
 -R_{mhk(i}R\du{j)rt}{m} \left( R^{kqsr}R\udu{t}{qs}{h}+
R^{ksqt}R\ud{hr}{qs}\right)\right. \nonumber\\
& - R_{kqs(i}R_{j)rmt}\left( R^{hsqt}R\ud{krm}{k}-R^{thsq}R\du{h}{rmk}\right)
 + \left( R_{itrj}R^{ksqt}R\udu{h}{sq}{r}\right)_{(;k;h)}
\lb{meo4}
\end{align}
$$\left. + \left( R_{isqt}R^{rktm}R\dud{j}{sq}{k}\right)_{(;r;m)}
-\left( R\ud{hrs}{(i} R_{j)mnr}R\du{h}{mnk} + R\ud{sht}{(i} 
R_{j)mnl}R\ud{kmn}{h}\right)_{(;k;s)}
\right] \nonumber $$

\section{Off-shell quartic terms in FRW cosmology}

The main Cosmological Principle of a {\it spatially\/} homogeneous and
isotropic $(1+3)$-dimensional universe (at large scales) gives rise to the
standard {\it Friedman-Robertson-Walker} (FRW) metric
\be
 ds^2 =  dt^2 - a^2(t)\left[ \fracmm{dr^2}{1-kr^2} +r^2d\Omega^2\right]
\lb{frw1} \ee
where the function $a(t)$ is known as the scale factor in  `cosmic' 
coordinates 
$(t,r,\theta,\phi)$; we use $c=1$ and $d\Omega^2= d\theta^2 
+\sin^2\theta d\phi^2$, 
while $k$ is the FRW topology index taking values $(-1,0,+1)$. Accordingly, 
the FRW 
metric (\ref{frw1}) admits a 6-dimensional isometry group $G$ that is either 
$SO(1,3)$, $E(3)$ or $SO(4)$, acting on the orbits $G/SO(3)$, with  the spatial
3-dimensional sections $H^3$, $E^3$ or $S^3$, respectively. By the coordinate 
change, 
$dt=a(t)d\eta$, the FRW metric (\ref{frw1}) can be rewritten to the form
\be ds^2 = a^2(\eta)\left[ d\eta^2 - \fracmm{dr^2}{1-kr^2}
 - r^2d\Omega^2\right]
\lb{frw2} \ee 
which is manifestly (4-dim) conformally flat in the case of $k=0$. Therefore, 
the 4-dim Weyl tensor of the FRW metric obvioulsy vanishes in the `flat' case 
of $k=0$. In fact, the FRW Weyl tensor vanishes in the other two 
cases, $k=-1$ and $k=+1$, too \cite{first}, thus 
\be C^{\rm FRW}_{ijkl}=0 \lb{frww}\ee

The inflation in early universe is defined as an epoch during which the scale 
factor is accelerating \cite{inf},
\be \ddt{a}(t)>0~,~{\rm or~ equivalently}~~
 \fracmm{d}{dt}\left( \fracmm{H^{-1}}{a}\right)<0 
\lb{idef}
\ee
where the dots denote time derivatives, and $H=\dt{a}/a$ is Hubble `constant'.
 The amount of inflation is given by a number of e-foldings \cite{inf},
\be N =\ln \fracmm{a(t_{\rm end})}{a(t_{\rm start})}= 
\int^{t_{\rm end}}_{t_{\rm start}}
 H~dt \lb{efol}\ee 
which should be around $70$ \cite{inf}. 

On the experimental side, it is known that the vacuum energy density 
$\r_{\rm inf}$ during inflation is bounded from above by a (non)observation of 
tensor fluctuations of the Cosmic Microwave Background (CMB) radiation 
\cite{cmbr}, 
\be \r_{\rm inf} \leq \left( 10^{-3}M_{\rm Pl}\right)^4 \ee     
It severly constrains but does not exclude the possibility of the geometrical 
inflation originating from the purely gravitational sector of string theory, 
because the factor of $10^{-3}$ above may be just due to some numerical 
coefficients.

Due to a single arbitrary function $a(t)$ in the FRW Ansatz (\ref{frw1}), it is
 enough to take only one gravitational equation of motion in eq.~(\ref{meo}) 
without matter, namely, its mixed $00$-component. As is well known \cite{inf}, 
the spatial (3-dimensional) curvature can be ignored in a very early universe, 
so we choose the manifestly conformally-flat FRW metric (\ref{frw1}) with 
$k=0$ in our Ansatz.  It leads to a purely gravitational equation of motion 
having the form 
\be \lb{4eqm} 3H^2\equiv
3\left(\fracmm{\dt{a}}{a}\right)^2=\b P_8\left( \fracmm{\dt{a}}{a},
\fracmm{\ddt{a}}{a}, \fracmm{\dddt{a}}{a},\fracmm{\ddddt{a}}{a}\right)~,
\ee
where $P_8$ is a {\it polynomial\/} with respect to its arguments, 
\be \lb{poly}
P_8= \sum_{n_1+2n_2+3n_3+4n_4=8,\atop n_1,n_2,n_3,n_4\geq 0} 
c_{n_1n_2n_3n_4} \left(\fracmm{\dt{a}}{a}\right)^{n_1}
\left(\fracmm{\ddt{a}}{a}\right)^{n_2}\left(\fracmm{\dddt{a}}{a}\right)^{n_3}
\left(\fracmm{\ddddt{a}}{a}\right)^{n_4} 
\ee
Here the sum goes over the {\it integer\/} partitions $(n_1,2n_2,3n_3,4n_4)$ of
 $8$, the dots stand for the derivatives with respect to time $t$, and 
$c_{n_1n_2n_3n_4}$ are some real coefficients. The highest derivative can enter
 only linearly, $n_4=0,1$.

The FRW Ansatz with $k=0$ yields the relevant curvatures as follows:
\be
R\ud{0}{i0j}=-\d_{ij}\ddt{a}a,~~ 
R\ud{i}{jkl}=\left(\d^i_k\d_{jl}-\d^i_l\d_{jk}\right)(\dt{a})^2,~~
R^{i}_{j}=\d^i_j\left[ \fracmm{\ddt{a}}{a}-2 \left(\fracmm{\dt{a}}{a}\right)^2
\right]
\lb{frwcurv}
\ee
where $i,j=1,2,3$.  For example, in the case of the
$(BR)^2$ gravity (\ref{meo4}), after a straightforward (though quite tedious) 
calculation of the mixed $00$-equation without matter and with the curvatures 
(\ref{frwcurv}), we find  
\begin{align}
3H^2 & + \beta\left[ 9\left( \fracmm{\ddt{a}}{a}\right)^4 -72 
H^2\left(\fracmm{\ddt{a}}{a}\right)^3 +96 H^4\left(\fracmm{\ddt{a}}{a}\right)^2
 -36H\left(\fracmm{\ddt{a}}{a}\right)^2
\left(\fracmm{\dddt{a}}{a}\right) \right.\nonumber\\
& \left. +75H^8  -72H^3\left(\fracmm{\ddt{a}}{a}\right)
\left(\fracmm{\dddt{a}}{a}
\right) + 24H^6\left(\fracmm{\ddt{a}}{a}\right)-24H^5
\left(\fracmm{\dddt{a}}{a}\right)\right]=0
\lb{br2frw}
\end{align}
It is remarkable that the 4th order time derivatives (present in various terms
 of eq.~(\ref{meo4})) cancel, whereas the square of the 3rd order time
derivative of the scale factor, $\dddt{a}{}^2$, does not appear at all in this 
equation.~\footnote{Taking Weyl tensors instead of Riemann curvatures leads to 
all vanishing coefficients.}

Our generalized Friedmann equation (\ref{4eqm}) applies to {\it any\/} 
combination of the quartic curvature terms in the action, including the 
Ricci-dependent terms. The coefficients $c_{n_1n_2n_3n_4}$ in eq.~(\ref{poly}) 
can be thought of as linear combinations of the coefficients in the most 
general quartic curvature action. The polynomial (6.3) has just about 10
 coefficients to be determined. 

Moreover, eqs.~(\ref{4eqm}) and (\ref{poly}) have the structure that
allows the existence of generic exact inflationary solutions without a 
spacetime singularity. Indeed, when using the most naive (de Sitter) Ansatz for
 the scale factor,       
\be  \lb{expf} a(t) =a_0 e^{At}  \ee
with some real positive constants $a_0$ and $A$, and substituting 
eq.~(\ref{expf}) into 
eq.~({\ref{4eqm}) we get $3A^2=(\#)\b A^8$, whose coefficient $(\#)$ is just 
a sum of all $c$-coefficients in eq.~(\ref{poly}). Assuming the  $(\#)$ to be 
positive, we find a solution
\be  \lb{simple}  A =  \left( \fracmm{3}{\#\b}\right)^{1/6} \ee

This solution in non-perturbative in $\beta$, i.e. it is impossible to get it 
when considering the quartic curvature terms as a perturbation. Of course, the 
assumption that we are dealing with the leading correction, implies 
$At\ll 1$. It leads to the natural hierarchy
\be \k M_{\rm KK}\ll 1 \quad {\rm or}\quad  l_{\rm Pl}\ll l_{\rm KK}\ee
where we have introduced the four-dimensional Planck scale $l_{\rm Pl}=\k$ and 
the compactification scale $l_{\rm KK}=M_{\rm KK}^{-1}$.

The exact solution (\ref{expf}) is non-singular, while it describes an 
inflationary isotropic and homogeneous early universe~\footnote{The exact de 
Sitter solutions in the special case (\ref{oco4}) were also found in 
ref.~\cite{ohta}.}. 

In the case of the quartic terms given by the Bel-Robinson tensor squared, i.e.
when ignoring all the Ricci tensor dependent terms in eq.~(\ref{oco4}), we find
that the coefficient $(\#)$ is negative, thus ruling out that option because
it does not allow the inflationary solution (\ref{simple}).

\section{Conclusion}

The higher curvature terms in the gravitational action defy the famous 
Hawking-Penrose theorem \cite{hawp} about the existence of a spacetime 
singularity in any exact solution to the Einstein equations. As we demonstrated
here, the initial cosmological singularity can be easily avoided by 
condsidering the superstrings-motivated higher curvature terms on 
equal footing (i.e. non-perturbatively) with the Einstein-Hilbert term.

As regards inflation, though we showed the natural existence of inflationary 
(de Sitter) exact solutions without a spacetime singularity under rather 
generic conditions on the coefficients in the higher-derivative terms, it is by
 no means sufficient, because our geometrical inflation is very short (not 
enough e-foldings), and has no end. In fact, we assumed the dominance of the 
higher curvature gravitational terms over all matter contributions in a very 
early universe at the Planck scale, and ignored all Kaluza-Klein modes beyond
four spacetime dimensions. Given the expansion of a four-dimensional universe
 under the geometrical inflation, the spacetime curvatures decrease, so that 
the mattter terms could no longer be ignored. The latter may effectively 
replace the geometrical inflation by another matter-dominated mechanism, 
allowing the inflation to continue substantially below the Planck scale.  
Needless to say, more research is needed in order to submit a specific 
mechanism for that.

The higher curvature terms are also relevant for the alternative (to inflation)
Brandenberger-Vafa scenario of string gas cosmology \cite{bvafa} ---
see e.g. ref.~\cite{borunda} for a recent (perturbative) investigation of the 
higher curvature corrections there.

The higher time derivatives in the cosmological equations are unavoidable with 
the higher curvature terms in the action, they should not be ignored (as e.g. 
in ref.~\cite{elizalde}), while they do not necessarily constitute a trouble.

The higher curvature terms may offer the alternative to the use of dynamical 
moduli, warped compactifications, or engineering desirable brane configurations
 \cite{binf,kklt,bbk}, since all those popular mechanisms need many 
{\it ad hoc} assumptions about a non-perturbative off-shell effective action 
of M-theory, while no explanation is provided why certain brane configurations 
are to be considered and how they arise.

Gravity with the quartic curvature terms is a good playground for going beyond 
the Einstein equations. Our analysis may be part of a more general approach 
based on superstrings, including moduli and extra dimensions.

\end{document}